\documentclass[%
 reprint,
 amsmath,amssymb,
 aps,prb,
 twocolumn,
 floatfix,
 showpacs]
 {revtex4-2}

\usepackage{graphicx}
\usepackage{dcolumn}
\usepackage{bm}
\usepackage{amsmath,amssymb}
\usepackage{hyperref}
\usepackage{natbib}
\usepackage{xcolor}
\usepackage[normalem]{ulem}
\usepackage{booktabs}
\usepackage{tabularx}
\usepackage{lipsum}
\usepackage{comment}
\usepackage{blindtext, xcolor}

\usepackage[font=small,format=plain,labelfont=bf,
singlelinecheck=false,justification=raggedright,
labelsep=space,figurename=FIG.]{caption}

\begin{document}

\preprint{APS/123-QED}

\title{Tuning magnetic interactions of Co and 4$d$ transition-metal atomic bilayers on Re(0001) via interface engineering} 

\author{Souvik Paul}
\email{souvikpaul@iisertvm.ac.in}
\affiliation{School of Physics, \\ Indian Institute of Science Education and Research Thiruvananthapuram, \\ Thiruvananthapuram, Kerala 695551, India}
\affiliation{Peter Gr$\ddot{u}$nberg Institut and Institute for Advanced Simulation, Forschungszentrum J$\ddot{u}$lich and JARA, 52425 J$\ddot{u}$lich, Germany.}

\author{Stefan Heinze}
\affiliation{Institute of Theoretical Physics and Astrophysics, Christian-Albrechts-Universit$\ddot{a}$t zu Kiel, Leibnizstrasse 15, 24098 Kiel, Germany}
\affiliation{Kiel Nano, Surface, and Interface Science (KiNSIS), University of Kiel, Germany}

\date{\today}

\begin{abstract}
	
Employing $\textit{ab-initio}$ density functional theory (DFT), we performed a systematic investigation of the electronic structures and the magnetic properties of atomic bilayers composed of a 4$d$ transition-metal layer (Rh, Pd and Ru) and a Co layer on the Re(0001) surface. Our DFT calculations reveal the influence of the bilayer composition and their stacking sequence on the magnetic ground state and magnetic interactions. We obtain the magnetic interactions by mapping the total energies onto an effective spin Hamiltonian which contains the exchange interaction and the Dzyaloshinskii-Moriya interaction (DMI) as well as the magnetocrystalline anisotropy energy (MAE). We observe noticeable changes in bilayer hybridization due to variation in bilayer composition and overlayer symmetry, leading to significant variation in magnetic interactions. In all considered systems, the nearest-neighbor exchange interaction is ferromagnetic, however, the value varies by up to a factor of 5. The effective, nearest-neighbor DMI constant exhibits variation in sign over the films considered, favoring either right- (clockwise) or left-handed (anticlockwise) cycloidal spin spirals. The value of the DMI changes by up to a factor of 27 among the films. For most of the systems, the MAE favors an out-of-plane easy magnetization axis, however, for hcp-Co/Rh and hcp-Co/Ru bilayers on Re(0001), it prefers an in-plane magnetization axis. The magnitude of the MAE varies from a small value of about 0.1 meV/Co atom up to about 2 meV/Co atom for Co/Pd bilayers. The spin spiral energy dispersion curve rises quite quickly close to the ferromagnetic state for films in which the Co layer is adjacent to the vacuum indicating a large effective exchange constant which stabilizes a ferromagnetic ground state in Co/$4d$ bilayers on Re(0001). The energy dispersion curve becomes flatter for films with a Co layer that is sandwiched between a $4d$ overlayer and the Re(0001) surface. In this case, the exchange constant is much reduced and the ground state is determined by the competition among the exchange interaction, favoring the FM state, the DMI, which favors cycloidal spin spirals, and the MAE, which disfavors spin spirals over the FM state. As a result, hcp-Rh/Co/Re(0001) shows a spin spiral ground state driven by DMI with a period of 13 nm while the other films exhibit a ferromagnetic ground state. The spin spiral energy dispersion of hcp-Rh/Co/Re(0001) indicates that isolated skyrmions can be stabilized in the ferromagnetic background with an applied magnetic field. Our results further suggest that isolated skyrmions could be realized even in absence of an external field in fcc-Rh/Co/Re(0001), hcp-Pd/Co/Re(0001), fcc-Pd/Co/Re(0001) and fcc-Ru/Co/Re(0001). This makes ultrathin films composed of a 4$d$ transition-metal and a Co layer on Re(0001) promising candidates for the search of isolated skyrmions.           
 
\end{abstract}

\maketitle


\section{Introduction}

Skyrmion lattices were experimentally discovered at low temperature in bulk systems \cite{bulk1,bulk3} and at surfaces \cite{heinze2011}. Subsequently, isolated magnetic skyrmions \cite{bogdanov1989} were observed in ultrathin films \cite{romming2013} and in magnetic multilayers \cite{moreau2016,woo2016}. Since their discovery, skyrmions have been envisioned for many applications \cite{fert2017,back2020} ranging from next-generation logic devices \cite{rt3,rt1} and race-track memories \cite{rt1} to quantum computing \cite{Psaroudaki2021,Psaroudaki2022} as well as unconventional computing such as probabilistic \cite{pinna2018}, neuromorphic \cite{grollier2020} and reservoir computers \cite{prychynenko2018}. From an application point of view,  transition-metal based ultrathin films and multilayers are particularly promising in search for novel skyrmion hosting systems with target properties since the magnetic interactions, which stabilize skyrmions, can be tuned via interface structure and composition \cite{Dupe2016,Jia_2018,Jia2020,Nickel2023b}. 

Transition-metal films on heavy metal surfaces serve as model systems to uncover the underlying microscopic mechanisms related to skyrmions. The interfacial Dzyaloshinskii-Moriya interaction (DMI) \cite{dmi1,dmi2}, arises from inversion symmetry breaking at the interface and large spin-orbit coupling due to the heavy metal. It was first observed in a Mn monolayer on the W(110) surface \cite{dmibode}. The discovery of a nanoskyrmion lattice in an Fe monlayer on Ir(111) \cite{heinze2011} marks the beginning of exploring skyrmions at the interface. Stabilization of isolated skyrmions at the interface results from the interplay of DMI along with the Heisenberg pairwise exchange interaction and magnetocrystalline anisotropy.      

In contrast to investigations of skyrmions in Fe-based transition-metal ultrathin films \cite{heinze2011,romming2013,polesya2014,D2014,simon2014,romming2015,hagemeister2015,muckel2021,rozsa2016}, considerably less work have been done to date on Co-based ultrathin films. Nonetheless, is it evident from the existing reports that the Co-based films can host complex magnetic structures including skyrmions \cite{boulle2016,perini2018,herve2018,mayer2019}. In pseudomorphycally grown thin films of a Co monolayer and an atomic bilayer of Pt/Co on the Ir(111) surface, N\'eel-type domain walls with a clockwise rotational sense were observed in spin-polarized scanning tunneling microscopy (SP-STM) measurements at low temperature \cite{perini2018}. The first reported Co-based ultrathin film which hosts isolated skyrmions is Co/Ru(0001) \cite{herve2018}. The individual skyrmions in this film system are (meta)stabilized at a low magnetic field of 150 mT with a diameter of nearly 50 nm due to a competition between the Heisenberg exchange interaction and DMI. Isolated zero-field skyrmions with diameters of about 5 nm have been observed in a Rh/Co atomic bilayer on the Ir(111) surface \cite{mayer2019}. Large lifetimes have been predicted for skyrmions in Rh/Co/Ir(111) due to the frustration of exchange interactions \cite{Goerzen2023}. Even antiskyrmions have been proposed to be extremely stable in this film system with calculated lifetimes \cite{Goerzen2023} similar to those obtained for skyrmions in Pd/Fe/Ir(111) \cite{hagemeister2015,Malottki2019,Goerzen2020}.

Recently, a variety of complex magnetic structures has been found in transition-metal monolayers and transition/heavy metal bilayers deposited on the Re(0001) surface \cite{spethmann2020,paul2020,li2020,nickel2023}. These observations underscore the potential of this substrate as a promising candidate for realizing novel magnetic structures including skyrmions, alongside the well-known Ir and Rh substrates. The first experimental realization of the row-wise antiferromagnetic state and the 3$Q$ state, predicted theoretically nearly 20 years ago \cite{Kurz2001}, occurs in a Mn monolayer on Re(0001) \cite{spethmann2020}. Subsequently, the 3Q state has also been observed in Pd/Mn and Rh/Mn bilayers on Re(0001) through SP-STM studies \cite{nickel2023}. Theoretical studies have demonstrated that the isolated skyrmions can be stabilized in a Pd/Fe bilayer on Re(0001) with the aid of an external magnetic field \cite{paul2020,paul2022}. Moreover, theoretical investigations have highlighted the importance of the complex noncollinear structures in triggering topological superconductivity when placed in proximity to a conventional superconductor \cite{bedow2020}. Experimentally, nanoscale skyrmions have been realized in Pd/Fe bilayers grown on a few atomic layers of Ir on the Re(0001) surface \cite{Kubetzka2020}.

\begin{figure}[!htbp]
	\includegraphics[scale=1.0]{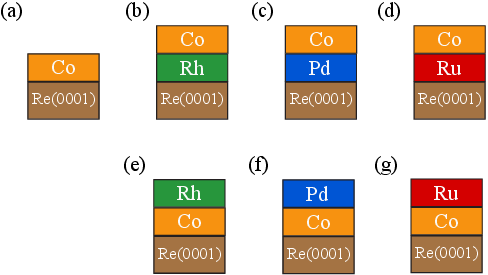}
	\centering
	\caption{Sketches of the ultrathin films considered in this study. Top row shows four film systems with a Co overlayer: (a) Co (b) Co/Rh, (c) Co/Pd, and (d) Co/Ru on Re(0001). Bottom row shows three films where the bilayer composition is swapped: (e) Rh/Co (f) Pd/Co and (g) Ru/Co on Re(0001). For each film, except for Co/Re(0001), hcp and fcc stacking orders of the overlayer are considered which adds up to thirteen films.} 
	\label{fig:f1}
\end{figure}

Here, we study the magnetic interactions in atomic bilayers, consist of a Co and a $4d$ transition metal (Ru, Rh, Pd) layers, on the Re(0001) surface using the $\textit{ab-initio}$ density functional theory (DFT) methods. We consider different compositions, i.e.,  a Co/$4d$ and a $4d$/Co bilayers, and fcc and hcp stacking of the overlayer (Fig. \ref{fig:f1}). As a reference system, we include a Co monolayer in hcp stacking on Re(0001). We perform DFT calculations for various collinear and noncollinear magnetic states including the effect of spin-orbit coupling (SOC). The goal of our investigation is to identify systems which are promising for the realization of topological spin structures such as skyrmions and antiskyrmions. 

Based on our DFT calculations, we show that the bilayer hybridization changes significantly with the variation of bilayer composition and symmetry of the overlayer resulting in a change of spin spiral energy dispersion across the films considered. We find a spin spiral ground state for hcp-Rh/Co/Re(0001) with a period of 13 nm, while all other films exhibit a ferromagnetic ground state. Interestingly, the energy dispersion indicates that isolated skyrmions can be found in hcp-Rh/Co/Re(0001) by applying an external magnetic field. The flat energy dispersion of spin spirals in the vicinity of the ferromagnetic ground suggests that zero-field magnetic skyrmions can be realized in four of the studied film systems: fcc-Rh/Co/Re(0001), hcp-Pd/Co/Re(0001), fcc-Pd/Co/Re(0001) and fcc-Ru/Co/Re(0001). 

We evaluate the exchange interaction and the DMI within the nearest-neighbor approximation by mapping the total DFT energies to an effective atomistic spin model. In addition, we calculate the magnetocrystalline anisotropy energy (MAE) to identify the easy magnetization direction of the films. The spin spiral dispersion of films with a Co overlayer is very steep and the effective nearest-neighbor exchange constant is quite large. For systems in which the Co layer is sandwiched between the $4d$ overlayer and the Re(0001) surface, the dispersion is flatter close to the ferromagnetic state and the exchange constant is much reduced. For selected film systems, we obtain exchange constants beyond nearest-neighbors which allows future spin dynamics simulations to verify the predicted occurrence of skyrmions. Given the experimental evidence of a submonolayer pseudomorphic growth of Co on Re(0001) \cite{ouazi2014}, our proposed films present an intriguing prospect for further experimental investigations.    

This paper is organized in three sections. In the first section (section \ref{sec:compdet}), we outline the methodology employed to compute the electronic structures and magnetic properties of ultrathin films and provide details of the calculations. In the second section (section \ref{sec:resdiss}), we present and analyze the results of all thirteen ultrathin films, as well as provide an in-depth discussion. In the final section, section \ref{sec:conc}, we offer a comprehensive summary of the investigation.

\section{\label{sec:compdet} Computational details}

To compute the electronic and magnetic structures of ultrathin films, we used two of the most accurate first-principles DFT methods, namely, the projected augmented wave (PAW) method as implemented in the $\textsc{vasp}$ code~\cite{vasp} and the full-potential linearized augmented plane-wave (FLAPW) method as implemented in the $\textsc{fleur}$ code~\cite{fleur}.  Various preceding reports have shown the accuracy and efficiency of these two DFT codes in computing the electronic structures and magnetic properties of ultrathin transition-metal films \cite{heinze2011,rhfeir111,ferrini2008,yoshida2012,paul2020}. 

We used the $\textsc{vasp}$ code to perform structural relaxations of the films in collinear magnetic states. In line with previous works \cite{paul2020,hoi2020,paul2022,li2020}, we have chosen the in-plane lattice constant ($a_{in}$) of the film as the experimental lattice constant of bulk Re, $a$= $2.761$ \AA, while the interlayer distance ($d_{0}$) of the films was set to the bulk layer separation  $c/a$= $2.228$ \AA \ \cite{re}. We modeled the experimental film geometry as an asymmetric film comprising of a monolayer or a bilayer on top of a nine layers Re(0001) substrate. To obtain the equilibrium geometry of these films, we allowed the top three layers (two for hcp-Co/Re(0001)) to relax along the $z$-direction via minimizing the forces on all atoms to less than $0.01$ eV/\AA\ using the $\textsc{vasp}$ code. We used the generalized gradient approximation (GGA) as parameterized by Perdew, Bruke and Ernzerhof \cite{ggapbe} for the exchange-correlation part of the potential. The energy cut-off was set to 500 eV and we used 66 $k$ points in the irreducible part of the two-dimensional Brillouin zone (2DBZ) to perform the BZ integration.

The magnetic properties of the structurally optimized films were obtained using the $\textsc{fleur}$ code. In particular, we calculated the energy dispersion $E(\mathbf{q})$ of homogeneous flat spin spirals self-consistently in the scalar-relativistic approximation \cite{kurz2004} as a function of wave vector $\mathbf{q}$. The magnetic moment $\textbf{M}_{i}$ at the atomic site $\textbf{R}_{i}$ of a spin spiral is specified by $\textbf{M}_{i}$=$M \left(\textrm{cos}(\textrm{\textbf{q}}
\cdot \textrm{\textbf{R}}_{i}), \textrm{sin}(\textrm{\textbf{q}} \cdot \textrm{\textbf{R}}_{i}), 0 \right)$, where $M$ is the magnitude of the moment. For spin spiral calculations, we used the local density approximation (LDA) of the exchange-correlation potential, as parameterized by Vosko, Wilk and Nusair \cite{vwn}. The total DFT energy can be mapped to the Heisenberg Hamiltonian

\begin{align} \label{eq1}
\mathcal{H}_{\rm ex} =- \sum_{ij} J_{ij} \ (\textbf{m}_{i}\cdot\textbf{m}_{j})
\end{align}

to evaluate the exchange interaction constants $J_{ij}$ between magnetic moments $\textbf{m}_{i}$ and $\textbf{m}_{j}$ at sites $i$ and $j$, respectively. We used the generalized Bloch theorem to compute the energy dispersion \cite{kurz2004} which allows us to use the chemical unit cell avoiding the large supercell calculations. A dense mesh of 48$\times$48 $k$-points in full 2DBZ was used to compute the full energy dispersion as well as around the $\overline{\Gamma}$ point, $\mid\textbf{q}\mid \le 0.1 \times \frac{2\pi}{a}$. To get a good convergence in hcp-Co/Re(0001), we used a $k$-point mesh of 120$\times$120. We set the cut-off for the basis functions, $k_{\rm max}$, to 4.0 a.u.$^{-1}$. 

The DMI arises due to the concerted effect of inversion symmetry breaking at the interface and SOC, which favors a certain sense of rotation (clockwise or anticlockwise). Within the Fert-Levy model \cite{fertlevy}, the DMI Hamiltonian can be expressed as

\begin{align} \label{eq2}
\mathcal{H}_{\rm DMI} =- \sum_{ij} \textbf{D}_{\textrm{eff}}\cdot(\textbf{m}_{i}\times\textbf{m}_{j})
\end{align}

where $\textbf{D}_{\textrm{eff}}$ is the effective DMI vector. Since the DMI energy is quite small compared to the total energy of the system, we evaluate it within the first-order perturbation theory based on the self-consistent spin spiral calculations \cite{heide,Zimmermann2014}. Due to the film's inherent symmetry, the DMI vectors lie in the plane of the interface and they are perpendicular to the line joining $\textbf{m}_{i}$ and $\textbf{m}_{j}$, thus, favoring cycloidal spin spirals. The positive (negative) sign of the DMI constant indicates that a right-rotating (left-rotating), i.e., clockwise (anticlockwise), spin spiral is preferred.

The second energy contribution arising from SOC is the MAE which was calculated self-consistently using the second variation approach \cite{li1990}. We performed fully self-consistent SOC calculations for the ferromagnetic state to obtain the energy difference between the out-of-plane ($\perp$) and in-plane ($\parallel$) magnetization directions, since MAE is defined as $K_{\mathrm{MAE}}$= $E_{\parallel}$ $-$ $E_{\perp}$. Therefore, a positive (negative) value of $K_{\mathrm{MAE}}$ signifies an out-of-plane (an in-plane) easy magnetization axis. To accurately calculate the in-plane and out-of-plane total energy difference, we used 13 to 17 substrate layers. To evaluate the MAE, we used the same $k$-point mesh and value of $k_{\rm max}$ as for the spin spiral calculations.

\begin{table*}[!htbp]
	\centering
	\caption{Structural relaxed parameters for a Co monolayer and for atomic Co/4$d$ and 4$d$/Co bilayers on Re(0001). The interlayer distances and magnetic moments of the three outermost layers at their relaxed geometry in the FM state are given. The calculations are done with the $\textsc{vasp}$ code using the GGA functional. The energy difference between the AFM and FM state ($E_\mathrm{{AFM}}-E_\mathrm{{FM}}$) is also shown.}
	\label{tab:table1}
	\begin{ruledtabular}
		\begin{tabular}{cccccccc} 
			Systems & \begin{tabular}[c]{@{}c@{}}$\Delta d_{1}$ (\%)\end{tabular} & \begin{tabular}[c]{@{}c@{}}$\Delta d_{2}$ (\%)\end{tabular} & \begin{tabular}[c]{@{}c@{}}$\Delta d_{3}$ (\%)\end{tabular} & \begin{tabular}[c]{@{}c@{}}$\mu_{1}$\\ $(\mu_{\mathrm{B}})$ \end{tabular} & \begin{tabular}[c]{@{}c@{}}$\mu_{2}$\\ $(\mu_{\mathrm{B}})$ \end{tabular} & \begin{tabular}[c]{@{}c@{}}$\mu_{3}$\\ $(\mu_{\mathrm{B}})$ \end{tabular} & \begin{tabular}[c]{@{}c@{}}$E_\mathrm{{AFM}}-E_\mathrm{{FM}}$\\ (meV/Fe)\end{tabular} \\
			\midrule
			hcp-Co/Re(0001) & \begin{tabular}[c]{@{}c@{}}$-$12.93 \end{tabular} 
			& \begin{tabular}[c]{@{}c@{}}0.97 \end{tabular}            & \begin{tabular}[c]{@{}c@{}}--- \end{tabular}            & \begin{tabular}[c]{@{}c@{}}1.28 \end{tabular}               & \begin{tabular}[c]{@{}c@{}}$-$0.06 \end{tabular}               & \begin{tabular}[c]{@{}c@{}}--- \end{tabular}            & \begin{tabular}[c]{@{}c@{}}110.74 \end{tabular} \\
			\midrule
			hcp-Co/Rh/Re(0001) & \begin{tabular}[c]{@{}c@{}}$-$9.34 \end{tabular} 
			& \begin{tabular}[c]{@{}c@{}}$-$3.05 \end{tabular}            & \begin{tabular}[c]{@{}c@{}}0.09 \end{tabular}            & \begin{tabular}[c]{@{}c@{}}1.93 \end{tabular}               & \begin{tabular}[c]{@{}c@{}}0.28 \end{tabular}               & \begin{tabular}[c]{@{}c@{}}$-$0.08 \end{tabular}            & \begin{tabular}[c]{@{}c@{}}112.09 \end{tabular} \\
			fcc-Co/Rh/Re(0001) & \begin{tabular}[c]{@{}c@{}}$-$7.99 \end{tabular} 
			& \begin{tabular}[c]{@{}c@{}}$-$1.71 \end{tabular}            & \begin{tabular}[c]{@{}c@{}}$-$0.81 \end{tabular}            & \begin{tabular}[c]{@{}c@{}}1.97 \end{tabular}               & \begin{tabular}[c]{@{}c@{}}0.25 \end{tabular}               & \begin{tabular}[c]{@{}c@{}}$-$0.08 \end{tabular}            & \begin{tabular}[c]{@{}c@{}}111.50 \end{tabular} \\
			\midrule
			hcp-Rh/Co/Re(0001) & \begin{tabular}[c]{@{}c@{}}$-$11.13 \end{tabular} 
			& \begin{tabular}[c]{@{}c@{}}$-$8.89 \end{tabular}            & \begin{tabular}[c]{@{}c@{}}0.54 \end{tabular}            & \begin{tabular}[c]{@{}c@{}}0.67 \end{tabular}               & \begin{tabular}[c]{@{}c@{}}1.57 \end{tabular}               & \begin{tabular}[c]{@{}c@{}}$-$0.08 \end{tabular}            & \begin{tabular}[c]{@{}c@{}}111.57 \end{tabular} \\
			fcc-Rh/Co/Re(0001) & \begin{tabular}[c]{@{}c@{}}$-$10.23 \end{tabular} 
			& \begin{tabular}[c]{@{}c@{}}$-$6.64 \end{tabular}            & \begin{tabular}[c]{@{}c@{}}$-$0.36 \end{tabular}            & \begin{tabular}[c]{@{}c@{}}0.73 \end{tabular}               & \begin{tabular}[c]{@{}c@{}}1.67 \end{tabular}               & \begin{tabular}[c]{@{}c@{}}$-$0.09 \end{tabular}            & \begin{tabular}[c]{@{}c@{}}111.47 \end{tabular} \\
			\midrule
			hcp-Co/Pd/Re(0001) & \begin{tabular}[c]{@{}c@{}}$-$6.30 \end{tabular} 
			& \begin{tabular}[c]{@{}c@{}}2.03 \end{tabular}            & \begin{tabular}[c]{@{}c@{}}$-$2.12 \end{tabular}            & \begin{tabular}[c]{@{}c@{}}1.99 \end{tabular}               & \begin{tabular}[c]{@{}c@{}}0.16 \end{tabular}               & \begin{tabular}[c]{@{}c@{}}$-$0.02 \end{tabular}            & \begin{tabular}[c]{@{}c@{}}212.78 \end{tabular} \\
			fcc-Co/Pd/Re(0001) & \begin{tabular}[c]{@{}c@{}}$-$5.93 \end{tabular} 
			& \begin{tabular}[c]{@{}c@{}}2.50 \end{tabular}            & \begin{tabular}[c]{@{}c@{}}$-$2.37 \end{tabular}            & \begin{tabular}[c]{@{}c@{}}1.99 \end{tabular}               & \begin{tabular}[c]{@{}c@{}}0.14 \end{tabular}               & \begin{tabular}[c]{@{}c@{}}$-$0.02 \end{tabular}            & \begin{tabular}[c]{@{}c@{}}208.02 \end{tabular} \\
			\midrule
			hcp-Pd/Co/Re(0001) & \begin{tabular}[c]{@{}c@{}}$-$7.09 \end{tabular} 
			& \begin{tabular}[c]{@{}c@{}}$-$10.23 \end{tabular}            & \begin{tabular}[c]{@{}c@{}}0.54 \end{tabular}            & \begin{tabular}[c]{@{}c@{}}0.25 \end{tabular}               & \begin{tabular}[c]{@{}c@{}}1.51 \end{tabular}               & \begin{tabular}[c]{@{}c@{}}$-$0.08 \end{tabular}            & \begin{tabular}[c]{@{}c@{}}69.04 \end{tabular} \\
			fcc-Pd/Co/Re(0001) & \begin{tabular}[c]{@{}c@{}}$-$5.79 \end{tabular} 
			& \begin{tabular}[c]{@{}c@{}}$-$10.31 \end{tabular}            & \begin{tabular}[c]{@{}c@{}}0.96 \end{tabular}            & \begin{tabular}[c]{@{}c@{}}0.21 \end{tabular}               & \begin{tabular}[c]{@{}c@{}}1.48 \end{tabular}               & \begin{tabular}[c]{@{}c@{}}$-$0.08 \end{tabular}            & \begin{tabular}[c]{@{}c@{}}55.22 \end{tabular} \\
			\midrule
			hcp-Co/Ru/Re(0001) & \begin{tabular}[c]{@{}c@{}}$-$12.48 \end{tabular} 
			& \begin{tabular}[c]{@{}c@{}}$-$4.40 \end{tabular}            & \begin{tabular}[c]{@{}c@{}}1.89 \end{tabular}            & \begin{tabular}[c]{@{}c@{}}1.76 \end{tabular}               & \begin{tabular}[c]{@{}c@{}}0.16 \end{tabular}               & \begin{tabular}[c]{@{}c@{}}$-$0.08 \end{tabular}            & \begin{tabular}[c]{@{}c@{}}90.26 \end{tabular} \\
			fcc-Co/Ru/Re(0001) & \begin{tabular}[c]{@{}c@{}}$-$9.34 \end{tabular} 
			& \begin{tabular}[c]{@{}c@{}}$-$3.50 \end{tabular}            & \begin{tabular}[c]{@{}c@{}}0.98 \end{tabular}            & \begin{tabular}[c]{@{}c@{}}1.88 \end{tabular}               & \begin{tabular}[c]{@{}c@{}}0.29 \end{tabular}               & \begin{tabular}[c]{@{}c@{}}$-$0.06 \end{tabular}            & \begin{tabular}[c]{@{}c@{}}138.91 \end{tabular} \\
			\midrule
			hcp-Ru/Co/Re(0001) & \begin{tabular}[c]{@{}c@{}}$-$12.93 \end{tabular} 
			& \begin{tabular}[c]{@{}c@{}}$-$8.44 \end{tabular}            & \begin{tabular}[c]{@{}c@{}}0.54 \end{tabular}            & \begin{tabular}[c]{@{}c@{}}0.34 \end{tabular}               & \begin{tabular}[c]{@{}c@{}}1.21 \end{tabular}               & \begin{tabular}[c]{@{}c@{}}$-$0.06 \end{tabular}            & \begin{tabular}[c]{@{}c@{}}55.72 \end{tabular} \\
			fcc-Ru/Co/Re(0001) & \begin{tabular}[c]{@{}c@{}}$-$11.58 \end{tabular} 
			& \begin{tabular}[c]{@{}c@{}}$-$6.19 \end{tabular}            & \begin{tabular}[c]{@{}c@{}}$-$0.35 \end{tabular}            & \begin{tabular}[c]{@{}c@{}}0.61 \end{tabular}               & \begin{tabular}[c]{@{}c@{}}1.49 \end{tabular}               & \begin{tabular}[c]{@{}c@{}}$-$0.03 \end{tabular}            & \begin{tabular}[c]{@{}c@{}}60.41 \end{tabular} \\
		\end{tabular}
	\end{ruledtabular}
\end{table*} 

\section{\label{sec:resdiss} Results and discussion} 
\subsection{\label{sec:colmag} Structural parameters and collinear magnetism}

We start by presenting our results of two collinear configurations, i.e., the ferromagnetic (FM) and the row-wise antiferromagnetic (AFM) state, for all thirteen films in Table \ref{tab:table1}, as computed by the $\textsc{vasp}$ code using the GGA exchange-correlation potential.

To obtain the collinear magnetic state of lowest energy, we first relax the topmost three layers for all twelve films except hcp-Co/Re(0001), where only the two topmost layers, i.e., the Co overlayer and the first substrate Re layer are relaxed, perpendicular to the plane in the FM and AFM configurations, keeping all the other substrate layers fixed. We calculate the energy difference $\Delta E$=$E_{\mathrm{AFM}}-E_{\mathrm{FM}}$ for all films from the total energy of the FM and AFM configurations evaluated at their respective relaxed geometry. 

The energy difference $\Delta E$ is positive for all films (last column of Table \ref{tab:table1}) which illustrates that all systems prefer FM alignment of spins. However, the value of $\Delta E$ varies significantly among the films from 55~meV/Co atom for fcc-Pd/Co/Re(0001) to 212 meV/Co~atom for hcp-Co/Pd/Re(0001), which indicates that the latter film is a stronger ferromagnet than the former one. We observe that the energy difference $\Delta E$ is quite small for Pd/Co and Ru/Co bilayers, i.e., when the Co layer is sandwiched between the 4$d$ transition metal and Re substrate, as compared to Co/Pd and Co/Ru bilayers, i.e., when Co is an overlayer. Interestingly, the energy difference $\Delta E$ is similar for Co/Rh and Rh/Co bilayers.   

The optimized interlayer distances and the magnetic moments of the topmost three layers (two for hcp-Co/Re(0001)) are listed for the FM state in Table \ref{tab:table1}. We define the relaxation of the interlayer distances relative to the bulk value as follows

\begin{gather}
\Delta d_{\mathrm{i}} (\%)= \left(\frac{d_{\mathrm{i,i+1}}-d_{0}}{d_{0}} \right) \times 100 \label{eq12}
\end{gather}

where $d_{\mathrm{i,i+1}}$ is the relaxed interlayer distance between layer $i$ and $i+1$, and $d_{0}$ is the bulk interlayer distance of Re, i.e., 2.228 \AA. 

The relaxation of the overlayer is inward for all films and the first relaxed interlayer distance ($\Delta d_{1}$) is fairly large as compared to the second ($\Delta d_{2}$) and third ($\Delta d_{3}$) ones. However, the hcp and fcc stacked Pd films, i.e., hcp-Pd/Co/Re(0001) and fcc-Pd/Co/Re(0001), behave in a different manner. The relaxed second interlayer distance ($\Delta d_{2}$) of the two Pd/Co/Re(0001) films is larger than the first one ($\Delta d_{1}$). $\Delta d_{1}$ varies from $-5.8\%$ for fcc-Pd/Co/Re(0001) to $-12.9\%$ for hcp-Co/Re(0001) and hcp-Ru/Co/Re(0001). We observe a variation of the first interlayer distance ($\Delta d_{1}$) by more than $3\%$ when the stacking order of the overlayer Co is changed from hcp to fcc in Co/Ru/Re(0001) films. 

The second interlayer relaxation is also inward for most films, except for both Co/Pd/Re(0001) films, which exhibit an outward relaxation. The value of the relaxed second interlayer distance ($\Delta d_{2}$) is significantly smaller for all films as compared to the first interlayer distance. An exception is the Pd/Co/Re(0001) films, where the second interlayer distance is considerably larger than the first one. 

The third interlayer spacing ($\Delta d_{3}$) is, in general, one order of magnitude smaller than the second one. However, for Co/Pd and Co/Ru bilayers on Re(0001), it is comparable to the second interlayer distance.  The three relaxed interlayer spacing indicate that they are strongly dependent on the composition of bilayers and the stacking symmetry of the overlayer. For example, a change in the stacking symmetry of the toplayer from hcp to fcc turns the third interlayer relaxation for Rh/Co and Ru/Co bilayers on Re(0001) from outward to inward. 

We notice a similar trend of relaxed interlayer distances in previously reported fcc and hcp stacked Rh overlayers on Co/Ir(111) \cite{mayer2019}. The relaxation of the topmost Rh and sandwiched Co layers are inward and the first relaxed interlayer distance ($\Delta d_{\textrm{Rh}-\textrm{Co}}$) is significantly larger than the second one ($\Delta d_{\textrm{Co}-\textrm{Ir}}$). The third interlayer relaxation ($\Delta d_{\textrm{Ir}-\textrm{Ir}}$) is outward and its magnitude is larger than the second interlayer distance. 

The magnetic moment of the Co overlayer varies only little for Co/Rh, Co/Pd and Co/Ru bilayers on Re(0001), i.e., between 1.8 $\mu_{\mathrm{B}}$ and 1.9 $\mu_{\mathrm{B}}$. It is reduced to 1.3~$\mu_{\mathrm{B}}$ for hcp-Co/Re(0001). The magnetic moment is also reduced for the Co layer, when it is sandwiched between the 4$d$ transition metal and Re substrate. It is about 1.6 to 1.7 $\mu_{\mathrm{B}}$ for Rh/Co bilayers, however, for fcc-Ru/Co on Re(0001), the value is much lower, i.e., 1.2 $\mu_{\mathrm{B}}$. The Co layer also induces a magnetic moment to the adjacent 4$d$ transition-metal layer and the first Re substrate layers which are aligned parallel and antiparallel to the Co moment, respectively. The value of the magnetic moment of 4$d$ transition-metal layer is one order of magnitude smaller than of the Co moment, whereas it is two order of magnitude smaller, i.e., nearly negligible, for the first Re substrate layer.

\subsection{\label{sec:noncolmag} Noncollinear magnetism and effective spin model}

In the preceding section, we examined the two collinear spin structures, namely, the FM and AFM states, and found that the lowest energy state of all the films is the FM spin configuration. In this section, we explore the existence of noncollinear ground state in these film systems taking the effect of SOC into account, i.e., energy contributions due to DMI and MAE. Below we present the energy dispersion $E(\mathbf{q})$ of homogeneous cycloidal spin spirals for all considered ultrathin films (Fig.~\ref{fig:f1}) as a function of wave vector $\textbf{q}$ including DMI and MAE calculated using the $\textsc{fleur}$ code in the vicinity of the $\overline{\Gamma}$ point ($\mid$$\textbf{q}$$\mid \leq 0.1\times\frac{2\pi}{a}$), which represents the FM state, along two high-symmetry directions $\overline{\Gamma \mathrm{KM}}$ and $\overline{\Gamma \mathrm{M}}$ of the 2DBZ.

\begin{figure}[!htbp]
	\includegraphics[scale=1.0]{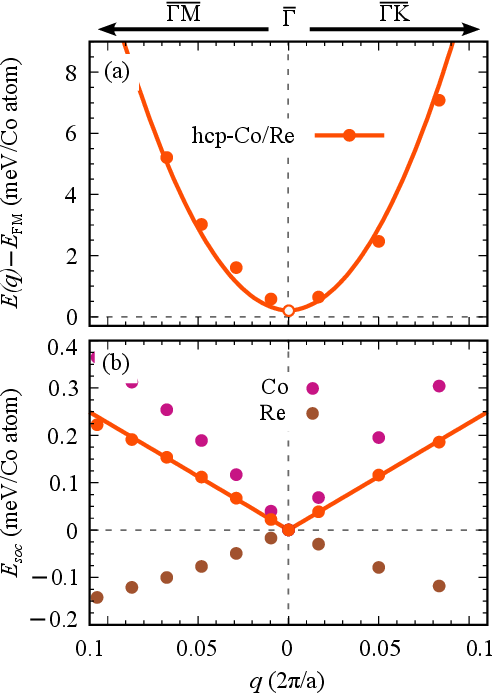}
	\centering
	\caption{(a) Energy dispersion $E$(\textbf{q}) of homogeneous left-rotating (anticlockwise) cycloidal spin spirals 
    for hcp-Co/Re(0001) in the vicinity of the $\overline{\Gamma}$ point (FM state) along the $\overline{\Gamma \mathrm{KM}}$ and $\overline{\Gamma \mathrm{M}}$ directions of 2DBZ. Filled circles present DFT data including SOC, i.e., DMI and MAE, and the solid lines are fit to the effective Heisenberg and DMI spin models. Note that the MAE shifts all spin spiral energies along the positive energy axis by $K_{\textrm{MAE}}$/2 with respect to the collinear FM state at zero energy. (b) Total energy contribution due to DMI (red) and individual contribution of the Co layer (purple) and the Re substrate (brown). Filled circles present DFT data and solid lines are fit to the effective DMI model (Eq. \ref{eq2}). The DFT calculations are done with the $\textsc{fleur}$ code.} 
	\label{fig:f2}
\end{figure}
 
 Conditions for observing skyrmions in an ultrathin film are either (i) it shows a shallow spin spiral energy minimum close to the FM state ($\overline{\Gamma}$ point) or (ii) it has a FM ground state, but exhibits a flat energy dispersion $E(\mathbf{q})$ close to the $\overline{\Gamma}$ point. In the first case, isolated skyrmions can be stabilized in the field-polarized phase by applying an external magnetic field at a minimum value that compensates the energy depth of the spin spiral minimum. An example for this case is the Pd/Fe bilayer on Ir(111) \cite{romming2013,D2014,Malottki2017a}. The second case is particularly intriguing since zero-field skyrmions can be stabilized under this condition in the ferromagnetic ground state. Recently, isolated skyrmions in the absence of an external field have been realized in Rh/Co bilayers on Ir(111) \cite{Meyer2019} meeting this condition and also a coexistence of nanoscale skyrmions and antiskyrmions has been predicted in this film \cite{Goerzen2023}.

For these reasons, we only display the energy dispersions up to a value of $q$ $\approx 0.1\times\frac{2\pi}{a}$ on both sides of the $\overline{\Gamma}$ point along $\overline{\Gamma \mathrm{KM}}$ and $\overline{\Gamma \mathrm{M}}$ directions in this section. We restrict the dispersion curve with a primary objective to identify the promising candidates among the thirteen films which can host skyrmions. In addition, our systematic study allows us to gain insight into the effect of bilayer composition and their associated symmetries on the effective magnetic interactions and hence on the magnetic ground state. In section \ref{sec:fulldisp}, we display the full dispersion curve for selected promising films and give a parameterization of an atomistic spin model containing beyond nearest-neighbour exchange constants required to accurately compute skyrmion-related properties using atomistic spin dynamics simulations \cite{Malottki2017a}.

Fig. \ref{fig:f2}(a) shows such an energy dispersion for the reference system of a Co monolayer in hcp stacking on the Re(0001) surface. The dispersion displays a parabolic shape, i.e., a $q^2$ dependence, close to the $\overline{\Gamma}$ point, characteristic of a ferromagnetic film. By fitting the DFT total energies obtained without SOC to Eq. \ref{eq1}, an effective nearest-neighbor exchange constant $J_{\rm eff}$ can be obtained (Table \ref{tab:table2}). In the limit of a long period, i.e., small value of $q$, a spin spiral state is energetically unfavorable than the FM state by $K_{\rm MAE}/2$ per spin due to the magnetocrystalline anisotropy energy (MAE). Therefore, the energy dispersion in Fig. \ref{fig:f2}(a) is shifted by $K_{\rm MAE}/2$.

Due to symmetry, the DMI favors cycloidal spin spirals in ultrathin films. The total DMI energy contribution to the dispersion of cycloidal spin spirals (Fig.~\ref{fig:f2}(b)) shows the characteristic linear dependence with $q$ in the vicinity of the $\overline{\Gamma}$ point. A fit in this linear regime allows to obtain the effective (nearest-neighbor) DMI constant, $D_{\rm eff}$.  The energy contribution due to DMI is quite small at $q$ $= 0.1\times\frac{2\pi}{a}$, i.e., only about 0.2 meV/Co atom, leading to a $D_{\rm eff}$ of $-$0.12 meV. The negative sign of $D_{\rm eff}$ (positive energy contribution in Fig. \ref{fig:f2}(b)) indicates a favorable anticlockwise rotational sense. In our DFT calculation, we can decompose the total energy contributions from DMI with respect to the different layers \cite{heide}. As seen in Fig. \ref{fig:f2}(b), the DMI energies due to Re and Co layers  exhibit opposite signs, i.e., opposite rotational sense, and the Co contribution dominates.

\begin{table}[!htbp]
	\centering
	\caption{Effective interaction constants for ultrathin films composed of Co/4$d$ and 4$d$/Co bilayers on Re(0001) (Fig. \ref{fig:f1}). The effective exchange ($J_{\mathrm{eff}}$) and the Dzyaloshinskii–Moriya ($D_{\mathrm{eff}}$) interactions constants are obtained from the fitting to total energy close to the $\overline{\Gamma}$ point ($\mid$$\textbf{q}$$\mid \leq 0.1\times\frac{2\pi}{a}$). $K_{\mathrm{MAE}}$ is calculated directly with the $\textsc{fleur}$ code. A positive (negative) DMI constant indicates that the right-rotating (left-rotating), i.e., clockwise (anticlockwise) spin spirals are favored. A negative (positive) $K_{\mathrm{MAE}}$ indicates an in-plane (out-of-plane) easy axis. The magnetic order of the ground state is also indicated. FM$^*$ indicates potential film for zero-field skyrmions.}
	\label{tab:table2}
	\begin{ruledtabular}
		\begin{tabular}{ccccc} 
			Systems 
			& \begin{tabular}[c]{@{}c@{}}$J_{\mathrm{eff}}$ \\ (meV)\end{tabular} 
			& \begin{tabular}[c]{@{}c@{}}$D_{\mathrm{eff}}$\\ (meV)\end{tabular}
			& \begin{tabular}[c]{@{}c@{}}$K_{\mathrm{MAE}}$\\ (meV)\end{tabular} 
			& \begin{tabular}[c]{@{}c@{}}Ground \\ state \end{tabular}\\
			\midrule
			hcp-Co/Re(0001) 
			& \begin{tabular}[c]{@{}c@{}}17.50 \end{tabular} 
			& \begin{tabular}[c]{@{}c@{}}$-$0.12 \end{tabular}
			& \begin{tabular}[c]{@{}c@{}}0.40 \end{tabular}            
			& \begin{tabular}[c]{@{}c@{}}FM \end{tabular}  \\
			\midrule
			hcp-Co/Rh/Re(0001)
			& \begin{tabular}[c]{@{}c@{}}25.70 \end{tabular} 
			& \begin{tabular}[c]{@{}c@{}}0.12 \end{tabular}
			& \begin{tabular}[c]{@{}c@{}}$-$1.16 \end{tabular}            
			& \begin{tabular}[c]{@{}c@{}}FM \end{tabular}  \\
			fcc-Co/Rh/Re(0001)
			& \begin{tabular}[c]{@{}c@{}}18.02 \end{tabular} 
			& \begin{tabular}[c]{@{}c@{}}0.16 \end{tabular}
			& \begin{tabular}[c]{@{}c@{}}0.54 \end{tabular}            
			& \begin{tabular}[c]{@{}c@{}}FM \end{tabular}  \\
			\midrule
			hcp-Rh/Co/Re(0001)
			& \begin{tabular}[c]{@{}c@{}}14.10 \end{tabular} 
			& \begin{tabular}[c]{@{}c@{}}1.91 \end{tabular}
			& \begin{tabular}[c]{@{}c@{}}0.10 \end{tabular}            
			& \begin{tabular}[c]{@{}c@{}}SS \end{tabular}  \\
			fcc-Rh/Co/Re(0001)
			& \begin{tabular}[c]{@{}c@{}}12.67 \end{tabular} 
			& \begin{tabular}[c]{@{}c@{}}1.17 \end{tabular}
			& \begin{tabular}[c]{@{}c@{}}0.60 \end{tabular}            
			& \begin{tabular}[c]{@{}c@{}}FM$^{*}$ \end{tabular}  \\
			\midrule
			hcp-Co/Pd/Re(0001)
			& \begin{tabular}[c]{@{}c@{}}27.40 \end{tabular} 
			& \begin{tabular}[c]{@{}c@{}}0.27 \end{tabular}
			& \begin{tabular}[c]{@{}c@{}}1.98 \end{tabular}            
			& \begin{tabular}[c]{@{}c@{}} FM \end{tabular}  \\
			fcc-Co/Pd/Re(0001)
			& \begin{tabular}[c]{@{}c@{}}26.51 \end{tabular} 
			& \begin{tabular}[c]{@{}c@{}}0.14 \end{tabular}
			& \begin{tabular}[c]{@{}c@{}}1.98 \end{tabular}            
			& \begin{tabular}[c]{@{}c@{}} FM \end{tabular}  \\
			\midrule
			hcp-Pd/Co/Re(0001)
			& \begin{tabular}[c]{@{}c@{}}15.12 \end{tabular} 
			& \begin{tabular}[c]{@{}c@{}}0.70 \end{tabular}
			& \begin{tabular}[c]{@{}c@{}}0.14 \end{tabular}            
			& \begin{tabular}[c]{@{}c@{}} FM$^{*}$ \end{tabular}  \\
			fcc-Pd/Co/Re(0001) 
			& \begin{tabular}[c]{@{}c@{}}5.47 \end{tabular} 
			& \begin{tabular}[c]{@{}c@{}}0.58 \end{tabular}
			& \begin{tabular}[c]{@{}c@{}}0.20 \end{tabular}            
			& \begin{tabular}[c]{@{}c@{}} FM$^{*}$ \end{tabular}  \\
			\midrule
			hcp-Co/Ru/Re(0001) 
			& \begin{tabular}[c]{@{}c@{}}13.73 \end{tabular} 
			& \begin{tabular}[c]{@{}c@{}}$-$0.32 \end{tabular}
			& \begin{tabular}[c]{@{}c@{}}$-$1.71 \end{tabular}            
			& \begin{tabular}[c]{@{}c@{}} FM \end{tabular}  \\
			fcc-Co/Ru/Re(0001) 
			& \begin{tabular}[c]{@{}c@{}}18.05 \end{tabular} 
			& \begin{tabular}[c]{@{}c@{}}0.07 \end{tabular}
			& \begin{tabular}[c]{@{}c@{}}1.23 \end{tabular}            
			& \begin{tabular}[c]{@{}c@{}} FM \end{tabular}  \\
			\midrule
			hcp-Ru/Co/Re(0001) 
			& \begin{tabular}[c]{@{}c@{}}18.10 \end{tabular} 
			& \begin{tabular}[c]{@{}c@{}}0.11 \end{tabular}
			& \begin{tabular}[c]{@{}c@{}}0.37 \end{tabular}            
			& \begin{tabular}[c]{@{}c@{}}FM \end{tabular}  \\
			fcc-Ru/Co/Re(0001)
			& \begin{tabular}[c]{@{}c@{}}6.90 \end{tabular} 
			& \begin{tabular}[c]{@{}c@{}}0.07 \end{tabular}
			& \begin{tabular}[c]{@{}c@{}}0.11 \end{tabular}            
			& \begin{tabular}[c]{@{}c@{}}FM$^{*}$ \end{tabular}  \\
		\end{tabular}
	\end{ruledtabular}
\end{table}

\begin{figure*}[!htbp]
	\includegraphics[scale=1.0]{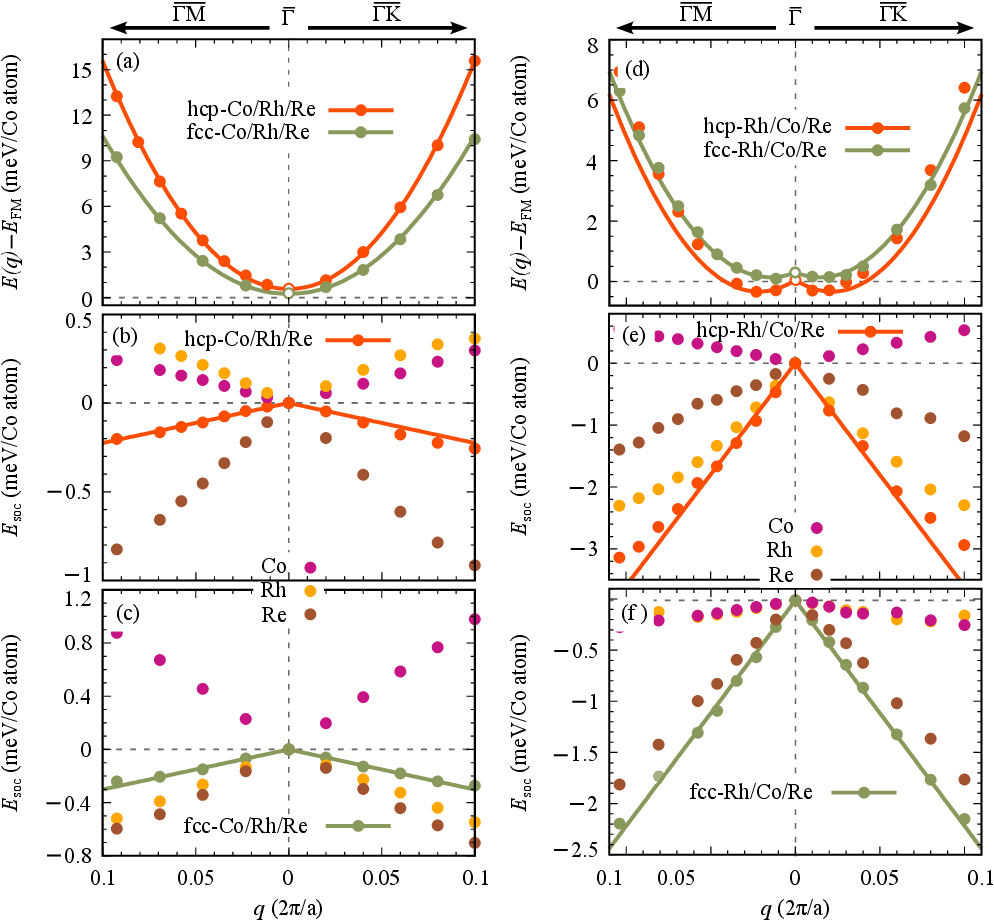}
	\centering
	\caption{Energy dispersion $E$(\textbf{q}) of homogeneous right-rotating (clockwise) cycloidal spin spirals around the $\overline{\Gamma}$ point along $\overline{\Gamma \mathrm{KM}}$ and $\overline{\Gamma \mathrm{M}}$ directions of 2DBZ for (a) hcp-Co and fcc-Co on Rh/Re(0001) and (d) hcp-Rh and fcc-Rh on Co/Re(0001). Filled circles present DFT data including SOC, i.e., DMI and MAE, and the solid lines are fit to the effective Heisenberg and DMI spin models. Note that the MAE shifts all spin spiral energies along the positive energy axis by $K_{\textrm{MAE}}$/2 with respect to the collinear FM state at zero energy. (b,c,e,f) Total energy contribution due to DMI (red for hcp-Co/Rh and hcp-Rh/Co and green for fcc-Co/Rh and fcc-Rh/Co) and individual contribution of the Co (purple ) and Rh (yellow) layers and the Re substrate (brown). Filled circles present DFT data and solid lines are fit to the effective DMI model (Eq. \ref{eq2}). The DFT calculations are done with the $\textsc{fleur}$ code.} 
	\label{fig:f3}
\end{figure*}

\begin{figure*}[!htbp]
	\includegraphics[scale=1.0]{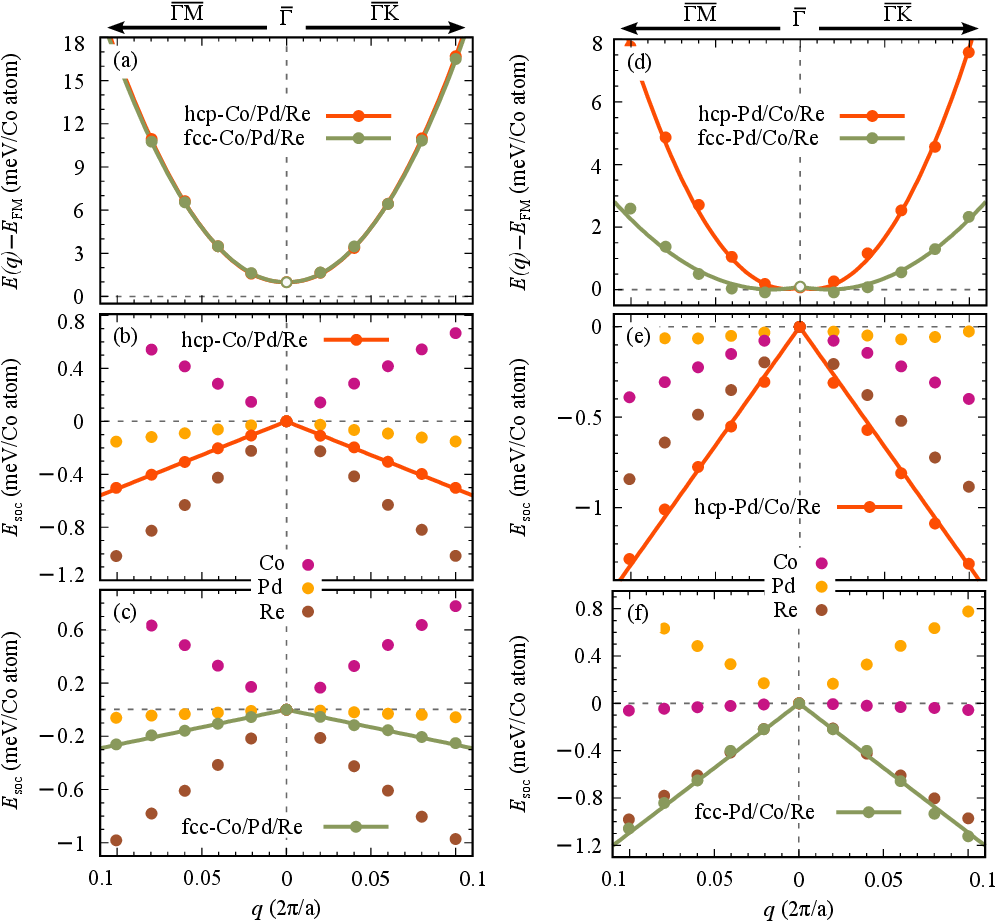}
	\centering
	\caption{Energy dispersion $E$(\textbf{q}) of homogeneous right-rotating (clockwise) cycloidal spin spirals around the $\overline{\Gamma}$ point along $\overline{\Gamma \mathrm{KM}}$ and $\overline{\Gamma \mathrm{M}}$ directions of 2DBZ for (a) hcp-Co and fcc-Co on Pd/Re(0001) and (d) hcp-Pd and fcc-Pd on Co/Re(0001). Filled circles present DFT data including SOC, i.e., DMI and MAE, and the solid lines are fit to the effective Heisenberg and DMI models. Note that the MAE shifts all spin spiral energies along the positive energy axis by $K_{\textrm{MAE}}$/2 with respect to the collinear FM state at zero energy. (b,c,e,f) Total energy contribution due to DMI (red for hcp-Co/Pd and hcp-Pd/Co and green for fcc-Co/Pd and fcc-Pd/Co) and individual contribution of the Co (purple) and Pd (yellow) layers and the Re substrate (brown). Filled circles present DFT data and solid lines are fit to the effective DMI model (Eq. \ref{eq2}). The DFT calculations are done with the $\textsc{fleur}$ code.} 
	\label{fig:f4}
\end{figure*}

\begin{figure*}[!htbp]
	\includegraphics[scale=1.0]{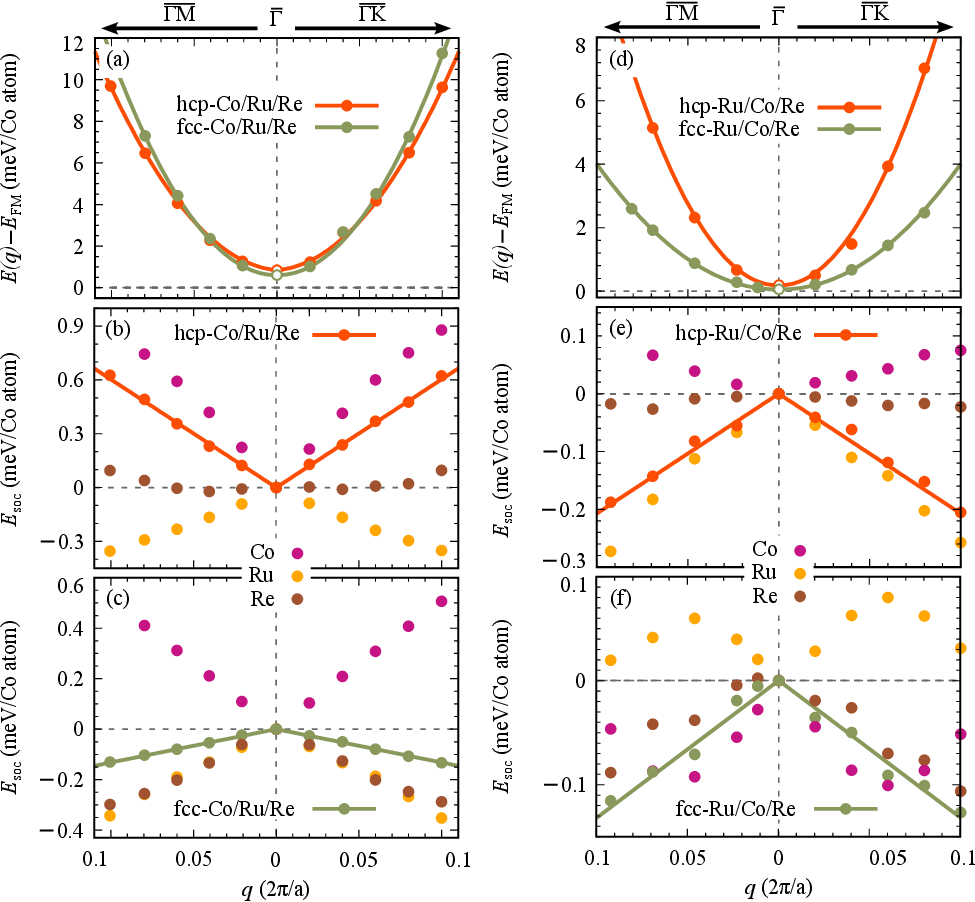}
	\centering
	\caption{Energy dispersion $E$(\textbf{q}) of homogeneous left-rotating (anticlockwise) cycloidal spin spirals for (a) hcp-Co/Ru/Re(0001) and right-rotating (clockwise) cycloidal spin spirals for (a) fcc-Co/Ru/Re(0001) and (d) hcp-Ru and fcc-Ru on Co/Re(0001) around the $\overline{\Gamma}$ point along $\overline{\Gamma \mathrm{KM}}$ and $\overline{\Gamma \mathrm{M}}$ directions of 2DBZ. Filled circles present DFT data including SOC, i.e., DMI and MAE, and the solid lines are fit to the effective Heisenberg and DMI models. Note that the MAE shifts all spin spiral energies along the positive energy axis by $K_{\textrm{MAE}}$/2 with respect to the collinear FM state at zero energy. (b,c,e,f) Total energy contribution due to DMI (red for hcp-Co/Ru and hcp-Ru/Co and green for fcc-Co/Ru and fcc-Ru/Co) and individual contribution of the Co (purple) and Ru (yellow) layers and the Re substrate (brown). Filled circles present DFT data and solid lines are fit to the effective DMI model (Eq.~\ref{eq2}). The DFT calculations are done with the $\textsc{fleur}$ code.} 
	\label{fig:f5}
\end{figure*}

Fig. \ref{fig:f3}(a,d) show the energy dispersion curves of spin spirals for Co/Rh and Rh/Co bilayers on Re(0001) for both fcc and hcp stacking of the topmost layer. Note that these dispersions include the effect of SOC, i.e., contribution of the DMI and the MAE, and display the energy of cycloidal spin spirals with a clockwise rotational sense favored by DMI.  We find that the curves for the films with a Co overlayer (Fig. \ref{fig:f3}(a)) are qualitatively similar to that for the Co monolayer on Re(0001) (Fig. \ref{fig:f2}(a)). For both stackings of the Co overlayer, we find a ferromagnetic ground state and a parabolic energy dispersion. The value of the effective exchange constant is considerably larger for the hcp-Co/Rh bilayer than for the fcc-Co/Rh bilayer or the Co monolayer on Re(0001) (Table \ref{tab:table2}). 

For the Rh/Co bilayers (Fig.~\ref{fig:f3}(d)), in which the Co layer is sandwiched between the Rh layer and the Re substrate, the energy dispersions look qualitatively different. Both curves display local energy minima for finite values of $q$ close to the $\overline{\Gamma}$ point. For hcp stacking of the Rh overlayer, a spin spiral energy minimum is obtained, while the ferromagnetic state (at zero energy) remains slightly favorable for fcc stacking.  The minimum of the spin spirals in hcp-Rh/Co/Re(0001) occurs at $q\approx$ 0.020 $\times\frac{2\pi}{a}$ with a pitch of $\lambda=$13.55 nm and it is nearly 0.35 meV/Co atom lower than the FM state. This shallow noncollinear minimum close to the FM state is similar to that observed in the prototypical skyrmion system Pd/Fe/Ir(111) \cite{Malottki2017a} and it implies that an external magnetic field of about 3 Tesla can lead to the field-polarized phase in which isolated skyrmions can be stabilized.

The lowest energy state of fcc-Rh/Co/Re(0001) (Fig.~\ref{fig:f3}(d)) is ferromagnetic. However, in contrast to the Co/Rh bilayers 
(Fig.~\ref{fig:f3}(a)), the dispersion curves are very flat in the vicinity of the $\overline{\Gamma}$ point and rise only gradually as $q$ increases. A similar type of flat dispersion curve has been reported for Rh/Co/Ir(111) \cite{Meyer2019} and it leads to the formation of isolated skyrmions in absence of an external magnetic field as observed using SP-STM studies \cite{Meyer2019,Perini2019}. Therefore, we anticipate that zero-field skyrmions could also occur in fcc-Rh/Co bilayer on Re(0001).

One reason for the different types of spin spiral energy dispersions obtained for Co/Rh versus Rh/Co bilayers on  Re(0001) is the considerably reduced effective exchange interaction for the latter films (Table \ref{tab:table2}). Another reason is the increase of the DMI constant by about one order of magnitude for the Rh/Co bilayers as compared Co/Rh bilayers on Re(0001) (Table \ref{tab:table2}). This effect is also clearly visible in the SOC contributions to the spin spiral dispersions (Figs. \ref{fig:f3}(b,c) versus (e,f)). Note that a linear fit to the total energy contribution due to SOC is excellent in the considered region close to the $\overline{\Gamma}$ point. 

The contributions from the Re substrate to the DMI energy is expected to be large since it is a heavy $5d$ transition metal with a large SOC constant. Interestingly, the Rh contribution is of similar magnitude as that of Re for fcc-Co/Rh/Re(0001) (Fig. \ref{fig:f3}(c)) and even exceeds the Re contribution for hcp-Rh/Co/Re(0001) (Fig. \ref{fig:f3}(e)). A detailed understanding of the contribution to the DMI from different atomic layers at the interface requires an in-depth analysis of the electronic structures which we shown in Ref. \cite{Nickel2023b} for a few systems.

For atomic bilayers consisting of a Co and a Pd layers, the general trend of the spin spiral energy dispersions (Fig. \ref{fig:f4}) is similar to that the above Rh bilayers. In particular, we find that for the films in which the Co layer is at the surface, i.e., Co/Pd/Re(0001) (Fig. \ref{fig:f4}(a)), $E(\mathbf{q})$ rises quite steeply in the vicinity of the $\overline{\Gamma}$ point, characteristic of a strongly ferromagnetic film. Accordingly, the values of $J_{\rm eff}$ (Table \ref{tab:table2}) are on the order of that for hcp-Co/Rh/Re(0001). In contrast, the Pd/Co bilayers on Re(0001) (Fig. \ref{fig:f4}(d)) display a rather flat energy dispersion which indicates significant frustration of exchange interactions. This is reflected in the relatively small effective nearest-neighbor exchange constants (Table \ref{tab:table2}). 

The DMI energy contribution to spin spirals is much enhanced for the sandwiched structures (Figs. \ref{fig:f4}(e,f)) compared to films with a Co overlayer (Figs. \ref{fig:f4}(b,c)). However, the effective DMI constant is smaller than that obtained for Rh/Co/Re(0001) (Table \ref{tab:table2}). Therefore, we do not find spin spiral energy minima for the Pd/Co bilayers on Re(0001), but observe a very shallow rise of the energy dispersion $E(\mathbf{q})$ (Fig. \ref{fig:f4}(d)). Concerning the possible formation of skyrmions, the ultrathin Pd/Co film with a fcc stacking of the Pd layer, i.e., fcc-Pd/Co/Re(0001) (Fig. \ref{fig:f4}(d)), is most promising due to the very slow rise of the energy of spin spirals with $\mathbf{q}$. On the other hand,
the energy dispersion rises more quickly close to the $\overline{\Gamma}$
point for hcp-Pd/Co/Re(0001), but the DMI is also increased (Table \ref{tab:table2}) as compared to fcc-Pd/Co/Re(0001). Since these films exhibits a ferromagnetic ground state and a flat energy dispersion, metastable, individual skyrmions might even appear without an applied magnetic field similar to observations in Rh/Co/Ir(111) \cite{Meyer2019}.

Finally, we turn to the bilayers consisting of a Ru and a Co layers on Re(0001) (Fig. \ref{fig:f5}). In these films, the overall trend is similar to that observed for the Pd or Rh bilayers (Figs. \ref{fig:f3} and \ref{fig:f4}, respectively). In particular, the energy dispersion rises quickly for Co/Ru/Re(0001) films (Fig. \ref{fig:f5}(a)) and the effective exchange constants are relatively large (Table \ref{tab:table2}). For the sandwich structured hcp-Ru/Co/Re(0001) (Fig. \ref{fig:f5}(d)), the energy dispersion rises almost as quickly as for the Co/Ru/Re(0001) film systems. But for fcc-Ru/Co/Re(0001) (Fig. \ref{fig:f5}(d)), we observe a much flatter energy dispersion with a small exchange constant $J_{\rm eff}$ of similar magnitude as for fcc-Pd/Co/Re(0001). However, the DMI, which favors clockwise cycloidal spin spirals in Ru/Co/Re(0001), is much smaller than the Co films with a Pd overlayer (Table \ref{tab:table2}). Therefore, the energy dispersion of spin spirals is not as shallow as for the fcc-Pd/Co bilayer. Thus we infer that the formation of isolated skyrmions in the ferromagnetic background might be possible for fcc-Ru/Co/Re(0001), but less favorable than the films with Pd/Co bilayers.

Table \ref{tab:table2} summarizes and quantifies the main results from the analysis of the spin spiral energy dispersions. The effective exchange $J_{\mathrm{eff}}$ and DMI $D_{\mathrm{eff}}$ constants for all thirteen films are listed, which are computed by fitting the spin spiral dispersions without and including SOC, respectively, in the vicinity of the $\overline{\Gamma}$ point within the nearest-neighbor approximation. In addition, the MAE and the spin structure of the ground state are given.

The effective exchange constant ($J_{\mathrm{eff}}$) provides valuable insights into the behavior of the dispersion curve within a narrow region around the $\overline{\Gamma}$ point. Precisely, it describes the curvature of the dispersion close to the $\overline{\Gamma}$ point. The positive value of the exchange constants indicates that the ground state of all the films is ferromagnetic when SOC was neglected. Moreover, the magnitude of the constant reflects the strength of the ferromagnetic interaction. Therefore, we can infer that hcp-Co/Pd/Re(0001) is the strongest ferromagnet and fcc-Pd/Co/Re(0001) is the weakest. We observe that the exchange constant of the films where Co is an overlayer is larger compared to the films with same composition but Co becomes a sandwiched layer. However, an exception is observed for hcp stacked Ru films. As the bilayer composition is flipped from hcp-Co/Ru to hcp-Ru/Co, instead of reducing, the effective exchange constant increases by 4.37 meV.   

The DMI energy is positive for hcp-Co/Re(0001) and hcp-Co/Ru/Re(0001) and negative for all other films. The positive (negative) value of DMI indicates that right-rotating (left-rotating), i.e., clockwise (anticlockwise) cycloidal spin spirals are favored. The magnitude of DMI constant is largest for hcp-Rh/Co/Re(0001), i.e., 1.91 meV and smallest for fcc-Co/Ru/Re(0001) and fcc-Co/Ru/Re(0001), i.e., 0.07 meV. 

Table \ref{tab:table2} also contains the MAE constants for all films under consideration. The magnitude of the MAE is larger for films with a Co overlayer as compared to films with a sandwiched Co layer due to the reduced coordination at the surface. The MAE attains a maximum value for hcp and fcc-Co/Pd/Re(0001) which is 1.98 meV/Co atom and the lowest value is 0.10 meV/Co atom obtained for hcp-Rh/Co/Re(0001). The MAE of hcp-Co/Rh/Re(0001) and hcp-Co/Ru/Re(0001) is negative and thus favors an in-plane easy magnetization direction. For all other films, an out-of-plane easy axis is obtained which favors the formation of skyrmions.

It is mentioned in the previous section that a large value of $J_{\mathrm{eff}}$ indicates a faster rise of $E(\mathbf{q})$ (without SOC) around the FM state ($\overline{\Gamma}$ point). As the value of $J_{\mathrm{eff}}$ is reduced, the dispersion curve becomes flatter. In such a situation, a significant DMI constant above 1 meV makes the spin spiral states more favorable than the FM state in hcp-Rh/Co and fcc-Rh/Co bilayers on Re(0001) (Fig.\ref{fig:f3}(d)). However, MAE acts opposite to the spin spiral stabilization, completely compensates the spin spiral energy of fcc-Rh/Co/Re(0001), leading to a FM ground state and partially compensates the spin spiral energy of hcp-Rh/Co/Re(0001), leading to a spin spiral ground state. Therefore, the stability of spin spiral states depends on the intricate interplay among these three interactions. 

\begin{figure*}[!htbp]
	\includegraphics[scale=1.0]{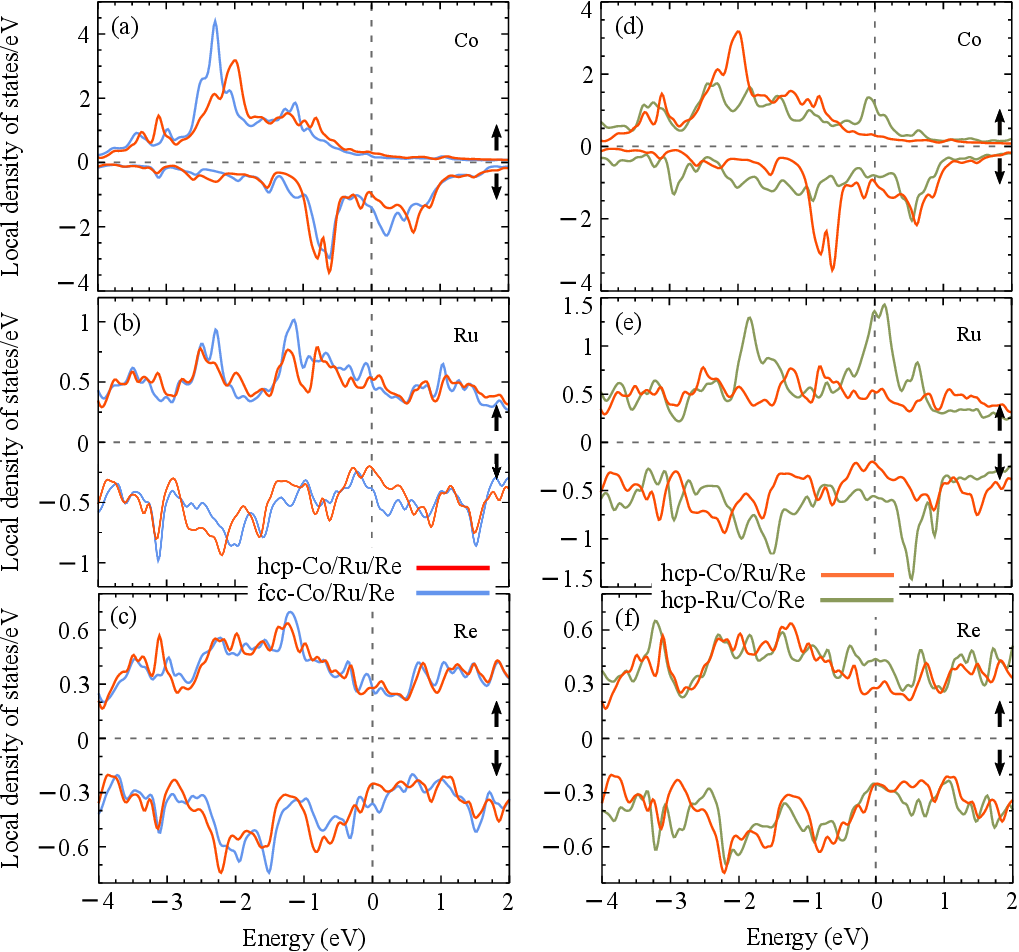}
	\centering
	\caption{Spin-polarized local density of states (LDOS) of the (a) Co, (b) Ru and (c) the first Re layers in hcp-Co/Ru/Re(0001) (red curves) and in fcc-Co/Ru/Re(0001) (blue curves). (d-f) Spin-resolved LDOS of the Co, Ru, and the first Re layers, respectively, in hcp-Co/Ru/Re(0001) (green curves) and in hcp-Ru/Co/Re(0001) (orange curves). The Fermi level is set to zero.} 
	\label{fig:f6}
\end{figure*}

Finally, a comparison with ultrathin Co-based films on the Ir(111) surface can be interesting. The effective exchange constant of hcp-Rh/Co and fcc-Rh/Co bilayers on Ir(111) is about 6.0 and 5.4 meV \cite{mayer2019}, respectively, i.e., nearly half of the value found for the Rh/Co/Re(0001) films. The reduction in the interaction constant leads to a flatter dispersion curve for Rh/Co/Ir(111) films around the $\overline{\Gamma}$ point compared to the Rh/Co/Re(0001) films. A sizable DMI constant of about 1.6 meV for hcp-Rh/Co/Ir(111) and about 0.8 meV for fcc-Rh/Co/Ir(111) favors right-rotating spin spirals over the FM state. Similar to the Rh/Co/Re(0001) films, due to a partial and complete compensation of the spin spirals energy by MAE, a FM and spin spiral ground states is observed in fcc-Rh/Co/Ir(111) and hcp-Rh/Co/Ir(111), respectively.

\subsection{\label{sec:elstr} Electronic structures}

In Table \ref{tab:table2}, we show the variation of the effective exchange interaction ($J_{\mathrm{eff}}$), the effective DMI ($D_{\mathrm{eff}}$) and the MAE ($K_{\mathrm{MAE}}$) constants across the films considered. The former two interaction constants vary as the hybridization of the Co layer changes with the adjacent $4d$ and $5d$ transition-metal layers, whereas the MAE constant ($K_{\mathrm{MAE}}$) is sensitive to changes of the density of states (DOS) at the Fermi level. In this section, we demonstrate the effect of changing the overlayer stacking order (fcc versus hcp) and swapping the bilayer composition on the local DOS (LDOS). To explore this, we choose three films: a hcp and fcc stacked Co monolayer on Ru/Re(0001) and the sandwiched structure hcp-Ru/Co/Re(0001), which is promising for observing skyrmions. We choose Ru films since the largest variation of LDOS is observed among these films. The change in LDOS of Co, Ru, and the first Re substrate layer for a variation of the overlayer stacking from hcp-Co to fcc-Co on Ru/Re(0001) are displayed in Fig. \ref{fig:f6}(a-c), respectively, and the LDOS of the three layers in the same order for swapping of the bilayer composition from hcp-Co/Ru to hcp-Ru/Co on Re(0001) are displayed in Figs. \ref{fig:f6}(d-f), respectively. The LDOS are evaluated in the FM state using the $\textsc{fleur}$ code.

Before pointing out the intricate details, we summarize the general features of LDOS. Since the films consist of 3$d$, 4$d$ and 5$d$ transition-metal elements, the LDOS are predominantly influenced by the $d$ orbitals. It is noteworthy that the $d$ orbitals become more delocalized as one moves from 3$d$ to 4$d$ and 5$d$ elements, i.e., from Co, to Ru and Re, respectively. As a result, the value of the Co LDOS is significantly higher than that of the other two elements and the intensity decreases as one moves from Ru to Re. Furthermore, the change in the LDOS of Co and Ru due to swapping of bilayer elements is quite substantial as compared to the change upon varying the overlayer stacking symmetry. This can be attributed to the reduced hybridization in case of a Co overlayer which is adjacent to only the Ru layer (Fig. \ref{fig:f6}(a)). In contrast, for the sandwiched structure hcp-Ru/Co/Re(0001), the Co layer is affected by both the Ru and the adjacent Re layers (Fig. \ref{fig:f6}(d)). A direct consequence of the increased hybridization is a much reduced magnetic moment of the sandwiched Co layer, i.e., 1.21 $\mu_{\rm B}$, as compared to a value of 1.76 $\mu_{\rm B}$ in case of a Co overlayer (see Table \ref{tab:table1}). The LDOS of Co show a large exchange splitting of majority and minority states and the induced spin-polarization on the Ru layer is clearly visible. The spin majority and minority LDOS of Re are nearly symmetric resulting in a small induced magnetic moment for all three films (Table \ref{tab:table1}).

We observe that as the stacking order of the overlayer Co varies from hcp-Co to fcc-Co on Ru/Re(0001) (Fig.~\ref{fig:f6}(a)), the LDOS of Co change notably. Conversely, the effect of the variation in overlayer stacking on the LDOS of Rh and the first Re layer is relatively modest (Fig. \ref{fig:f6}(b,c)). The spin majority peak of Co at $-2$ eV for hcp-Co/Ru/Re(0001) shifts to $-2.3$ eV for fcc-Co/Ru/Re(0001). The intensity of the latter peak increases and becomes sharper than the former one. The spin minority peak of Co for hcp-Co/Ru/Re(0001) at 0.6 eV above the Fermi level moves closer to the Fermi level as the stacking of Co changes to fcc, and the intensity increases by a little amount.

There is almost no effect on the LDOS of Re (Fig. \ref{fig:f6}(f)) upon changing the bilayer composition from Co/Ru to Ru/Co. On the other hand, significant differences are observed in the LDOS of Co (Fig. \ref{fig:f6}(d)) and Ru (Fig. \ref{fig:f6}(e)). The spin majority LDOS of Co for the overlayer and sandwiched layers are quite small above 1 eV. Below this energy, the LDOS for the sandwiched Co layer increase faster than that of the Co overlayer, become higher in magnitude and show a peak at 0.7 eV below the Fermi level. Thereafter, the intensity of the LDOS of the Co overlayer increases and becomes comparable to the sandwiched layer from the energy range of $-0.50$ eV to $-4$ eV, except around $-2$ eV. Around that energy, the LDOS of the Co overlayer show a sharp peak while the LDOS from the sandwich layer exhibit a valley. The spin minority LDOS of Co for both layer positions are similar above 0.5 eV and follow each other. Below that energy, the differences become prominent. The LDOS of the Co overlayer show a small peak at 0.25 eV and a large double peak around 0.75 eV below the Fermi level, while the intensity of the LDOS of the Co sandwich layer remains fairly small. Below $-1$ eV, the intensity of the LDOS from Co overlayer reduces and becomes lower than the LDOS of the sandwiched Co layer. 

The intensity of spin majority LDOS of Ru for hcp-Ru/Co/Re(0001) remains nearly same over the energy range, but for hcp-Co/Ru, it shows a two-peak structure with a significantly higher intensity as compared to hcp-Ru/Co/Re(0001). The peaks are located at the Fermi level and 0.75 eV below the Fermi level. Beside these variations, the intensity of LDOS for these two films are comparable. In contrast, the spin majority LDOS are quite different. The intensity of LDOS originating from Ru overlayer is higher above $-2$ eV and below that energy, it is comparable to the sandwiched Ru layer. We observe a sharp peak in the LDOS from Ru overlayer about 0.25 eV accompanied by a small peak below 1.5 eV of the Fermi level.
 
An comparison of the LDOS of the Co/Ru versus Ru/Co bilayers on Re(0001) shows that there are large variations in the electronic structures. In particular, changing the position of the Co layer from being the topmost layer to the sandwiched geometry results in a much stronger hybridization due to the interaction with the Ru and the Re layers. This leads to a reduced magnetic moment and a large modification of the interaction constants, especially a smaller effective exchange constant reflecting exchange frustration (Table \ref{tab:table2}). In contrast, changing the stacking order of the overlayer from hcp to fcc affects the LDOS less in accordance with the qualitatively similar spin spiral energy dispersion (Fig. \ref{fig:f5}).

\subsection{\label{sec:fulldisp} Full energy dispersion and beyond nearest-neighbor exchange interactions}

\begin{table*}[!thbp]
	\centering
	\caption{Beyond nearest-neighbor exchange interactions. i$^{th}$ nearest-neighbor exchange interaction constants, $J_{i}$, are presented for three selected films, hcp-Rh/Co/Re(0001), fcc-Rh/Co/Re(0001), and fcc-Pd/Co/Re(0001) which are promising to host skyrmions and for hcp-Co/Re(0001) as a reference system. The exchange constants are given in meV.} 
	\label{tab:table3}
	\begin{ruledtabular}
		\begin{tabular}{ccccccccccc}
			Systems & $J_{1}$ & $J_{2}$ & $J_{3}$ & $J_{4}$ & $J_{5}$ & $J_{6}$ & $J_{7}$ & $J_{8}$ & $J_{9}$ & $J_{10}$ \\
			\colrule
			hcp-Co/Re(0001) &9.18 &12.43 &$-$0.88 &$-$1.65 &$-$0.03 &0.92 &$-$0.92 &1.04 &0.19 &$-$0.24 \\
			hcp-Rh/Co/Re(0001) &16.64 &0.80 &$-$0.21 &$-$0.19 &$-$0.01 &$-$0.27 &0.12 &$-$0.03 &$-$0.05 &0.08 \\
			fcc-Rh/Co/Re(0001) &15.72 &0.43 &0.25 &0.03 &$-$0.42 &--- &--- &--- &--- &---  \\
			fcc-Pd/Co/Re(0001) &7.04 &0.40 &$-$0.32 &0.12 &$-$0.46 &--- &--- &--- &--- &---
		\end{tabular}
	\end{ruledtabular}
\end{table*}

\begin{figure}[!htbp]
	\includegraphics[scale=1.0]{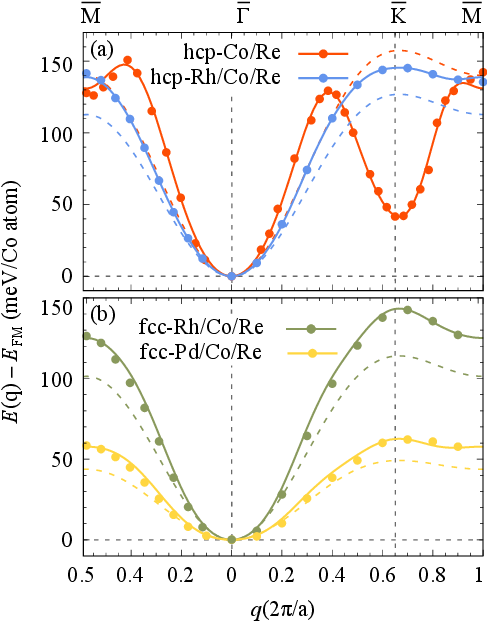}
	\centering
	\caption{Energy dispersion $E$(\textbf{q}) of homogeneous flat spin spirals without SOC along two high-symmetry directions $\overline{\Gamma \mathrm{KM}}$ and $\overline{\Gamma \mathrm{M}}$ of the 2DBZ for 
    (a) hcp-Co/Re(0001) (red) and hcp-Rh/Co/Re(0001) (blue), (b) fcc-Rh/Co/Re(0001) (green) and fcc-Pd/Co/Re(0001) (yellow). The solid circles present DFT data and the solid lines are fit to the beyond nearest-neighbor Heisenberg model (Eq. \ref{eq1}). The dashed lines are the fit within the nearest-neighbor Heisenberg model (effective spin model) discussed in section \ref{sec:noncolmag}} 
	\label{fig:f7}
\end{figure}

In section \ref{sec:noncolmag}, we displayed the energy dispersion of all Co/4$d$ and 4$d$/Co bilayers on Re(0001) from the $\overline{\Gamma}$ point to $\mid$$\textbf{q}$$\mid \leq 0.1\times\frac{2\pi}{a}$ and identified five 4$d$/Co bilayer films based on the characteristics of the dispersion curve as promising for skyrmions. We further computed the magnetic interactions 
and parameterized an effective spin model to get an insight into the variation of energy dispersion curves.

The effective spin model works well in the vicinity of the FM state, allowing us to select promising films without calculating the full dispersion curve, which is computationally expensive. However, as the $q$ value increases, the effective model starts to deviate from the DFT result as shown in Fig. \ref{fig:f7}, and prompts the necessity to incorporate the beyond nearest-neighbor interactions into the analysis. The beyond nearest-neighbor exchange terms do not contradict the qualitative conclusion drawn from the effective spin model on skyrmion formation. However, the beyond nearest-neighbor exchange interactions modifies skyrmion properties quantitatively such as critical magnetic fields, phase diagrams, radius, energy barriers and lifetime \cite{Malottki2017a,Malottki2019}. 

In Fig. \ref{fig:f7}, we display the full energy dispersion of homogeneous spin spirals as a function of wave vector $\textbf{q}$ excluding the effects of SOC along two high-symmetry directions $\overline{\Gamma \mathrm{KM}}$ and $\overline{\Gamma \mathrm{M}}$ of the 2DBZ for selected films:
hcp-Rh/Co (Fig. \ref{fig:f7}(a)), fcc-Rh/Co and fcc-Pd/Co (Fig. \ref{fig:f7}(b)) on Re(0001), which have potential for skyrmion formation and hcp-Co/Re(0001) (Fig. \ref{fig:f7}(a)), which is the reference system and not promising for skyrmions. The high symmetry points of the 2DBZ represent special spin structures: the FM state at the $\overline{\Gamma}$ point, row-wise AFM state at the $\overline{\mathrm {M}}$ point, and the N\'eel state at the $\overline{\mathrm K}$ point.

The energy dispersion curves $E(\mathbf{q})$ of hcp-Rh, fcc-Rh and fcc-Pd on Co/Re(0001) are qualitatively similar. The spin spiral energy increases as one moves from the $\overline{\Gamma}$ point along $\overline{\Gamma \mathrm{KM}}$ and attains a maximum value at the $\overline{\mathrm {K}}$ point and then decrease gradually with a further increase of $q$. On the other hand, the curves rise from the $\overline{\Gamma}$ point along $\overline{\Gamma \mathrm{M}}$ reache a maximum value at the $\overline{\mathrm {M}}$ point. The steepness of the dispersion curve of hcp-Rh and fcc-Rh on Co/Re(0001) is comparable, while it is quite low for fcc-Pd/Co/Re(0001). The energy difference between the AFM ($\overline{\mathrm {M}}$ point) and FM ($\overline{\Gamma}$ point) state is nearly 120 meV for hcp-Rh and fcc-Rh on Co/Re(0001), whereas the value is reduced by a factor of 2 to $\approx$ 60 meV in fcc-Pd/Co/Re(0001). Note that the dispersion curve of Rh/Co/Ir(111) which hosts zero-field skyrmions \cite{mayer2019} behaves similar to that of Rh/Co/Re(0001). The energy of the AFM state with respect to the FM state for Rh/Co/Ir(111) films is nearly 200 meV.

The reference system hcp-Co/Re(0001) shows a quite different behavior. The dispersion profile increases sharply from the $\overline{\Gamma}$ point along the $\overline{\Gamma \mathrm{KM}}$ and $\overline{\Gamma \mathrm{M}}$ directions and attains a maximum value around $q$$\approx 0.4\times\frac{2\pi}{a}$ for both directions. Then the curve gradually decreases with $q$ along $\overline{\Gamma \mathrm{M}}$ and obtains a value of 130 meV at the $\overline{\mathrm{M}}$ point. On the other hand, the curve decreases sharply after $q$$\approx 0.4\times\frac{2\pi}{a}$ along $\overline{\Gamma \mathrm{KM}}$ and exhibits a local minimum at the $\overline{\mathrm K}$ point. Again it increases sharply with $q$ and obtains a value of 130 meV at the $\overline{\mathrm{M}}$ point. In contrast to the other three curves, the maximum of this curve is obtained along the $\overline{\Gamma \mathrm{M}}$ direction.

To check the validity of the nearest-neighbor effective spin model, we display the full energy dispersion curve for all four films obtained by only considering the effective (nearest-neighbor) exchange constant $J_{\rm eff}$ listed in Table \ref{tab:table2} (dashed lines in Fig. \ref{fig:f7}). It is obvious that the effective model fits the DFT data very well close to the $\overline{\Gamma}$ point where the variation is parabolic ($q^2$) in nature. As the $q$ value increases, the curves start to deviate from a parabola, indicating a departure from the effective model and underlining the need to include interactions beyond nearest neighbors. To get insight into these exchange interactions, we fitted the full spin spiral dispersion to the Heisenberg model (Eq. \ref{eq1}). The obtained exchange constants for hcp-Co/Re(0001), hcp-Rh/Co/Re(0001), fcc-Rh/Co/Re(0001) and fcc-Pd/Co/Re(0001) are given in Table \ref{tab:table3}.

A comparison of the effective ($J_{\mathrm{eff}}$) and first ($J_{1}$) exchange interaction constants reveals that $J_{1}$ is slightly larger for the Rh/Co and Pd/Co bilayer on Re(0001). Since $J_{1}$ is approximately proportional to the energy difference between the FM and AFM state, this quantity is underestimated in the effective spin model. The larger value of $J_{1}$ is balanced by beyond nearest-neighbor exchange constants of negative sign which favor an antiparallel alignment of spins. Such an exchange frustration explains the flat energy dispersions in the vicinity of the $\overline{\Gamma}$ point (Fig. \ref{fig:f3}(d), Fig. \ref{fig:f4}(d) and Fig. \ref{fig:f5}(d)). Taking this effect into account is also crucial in terms of skyrmion stability since it has been shown that the energy barrier between a metastable skyrmion and the FM state is greatly enhanced by exchange
frustration \cite{Malottki2017a}. In addition, new collapse mechanisms of skyrmions such as the Chimera collapse can occur as shown in Pd/Fe/Ir(111) \cite{muckel2021} and in Rh/Co/Ir(111) \cite{Meyer2019}.

\section{\label{sec:conc} Conclusion}

We have investigated the possibility of complex noncollinear magnetic ground states in atomic bilayers consisting of a Co and a 4$d$ transition-metal layers on the Re(0001) surface using DFT. Our study encompasses three 4$d$ transition metals: Rh, Pd and Ru, which give rise to the combinations Co/Rh, Co/Pd, Co/Ru, Rh/Co, Pd/Co and Ru/Co. We have also considered hcp and fcc stacking of the topmost layer resulting in twelve ultrathin films. We have included a Co monolayer in hcp stacking on Re(0001) as a reference system. It is worth noting that pseudomorphic submonolayer growth of Co has been experimentally observed on Re(0001).

We have performed total energy calculations for the ferromagnetic and antiferromagnetic states on their respective relaxed geometries to obtain the lowest energy collinear spin structure. Further we calculated the energy dispersion of spin spirals via DFT including spin-orbit coupling within first-order perturbation theory. We obtained the effective nearest-neighbor exchange and DMI constants by fitting the spin spiral energy dispersions without and with SOC, respectively, in the vicinity of the FM state. The MAE constant is directly computed using DFT based on the energy difference between the in-plane and out-of-place magnetization directions in the FM state. Our results reveal that the magnetic interactions strongly depend on the composition and symmetry of the bilayer due to the hybridization of the Co layer with the $4d$ transition-metal layer and the Re substrate. 

Bilayers with a Co overlayer show a notably higher effective exchange constant than films with a sandwiched Co layer, indicating a steeper rise of the energy of spin spirals around the FM state for the former set of films. For films with a Co overlayer, the DMI constant is not large enough to stabilize a spin spiral state and the FM states remains the ground state. The curvature of the dispersion curve becomes flatter as the exchange interaction constant reduces for hcp-Rh/Co/Re(0001) and fcc-Rh/Co/Re(0001), and a higher DMI constant promotes right-rotating cycloidal spin spirals over the FM state. However, the MAE counteracts the spin spiral energy in fcc-Rh/Co/Re(0001) and stabilizes a FM ground state. In hcp-Rh/Co/Re(0001), the MAE can not completely compensate the spin spiral energy, resulting in a spin spiral ground state in this film with a period of about 13 nm. The dispersion curve becomes even flatter as the exchange constant further reduces for hcp-Pd/Co/Re(0001), fcc-Pd/Co/Re(0001) and fcc-Ru/Co/Re(0001) and an interplay between the DMI and MAE exhibit a FM ground state.  
 
The dispersion profile of fcc-Rh/Co/Re(0001), hcp-Pd/Co/Re(0001), fcc-Pd/Co/Re(0001) and fcc-Ru/Co/Re(0001) reveals that these films have potential to host isolated skyrmions at zero external magnetic field, whereas in case of hcp-Rh/Co/Re(0001), it necessitates an applied magnetic field of approximately 3 Tesla.

In conclusion, our first-principles DFT calculations show that atomic Rh/Co, Pd/Co and Ru/Co bilayers on the Re(0001) surface are extremely promising films in terms of observing magnetic skyrmions at finite and zero applied magnetic fields. Realizing zero-field magnetic skyrmions in ultrathin films on a surface which becomes superconducting at low temperature such as Re(0001) is particularly intriguing. Experimentally, nanoscale magnetic skyrmions have been realized in ultrathin Pd/Fe films on the Re(0001) surface, however, a significant external magnetic field is required \cite{Kubetzka2020}.

\section{\label{sec:ackn} Acknowledgments}

We gratefully acknowledge the North-German Supercomputing Alliance for providing the computing time. S.P. graciously acknowledges the support from IISER Thiruvananthapuram for funding and computing time at the Padmanabha cluster. S.P. also acknowledges the computing time from CDAC, India under the project ``topomagnetic-acc". S.P. further acknowledges financial support from the European Research Council (ERC) under the European Union’s Horizon 2020 research and innovation program (Grant No. 856538, project ``3D MAGiC". 

\end{document}